\documentclass[aps,prl,twocolumn,groupedaddress]{revtex4-1}

\usepackage{graphicx}
\usepackage{amsmath,amssymb}
\usepackage{bm}

\begin{document}


\title{Laser-induced Kondo effect in ultracold alkaline-earth fermions}


\author{Masaya Nakagawa}
\email[]{m.nakagawa@scphys.kyoto-u.ac.jp}
\author{Norio Kawakami}
\affiliation{Department of Physics, Kyoto University, Kyoto 606-8502, Japan}


\date{\today}

\begin{abstract}
We demonstrate that laser excitations can coherently induce a novel Kondo effect in ultracold atoms in optical lattices. Using a model of alkaline-earth fermions with two orbitals, it is shown that the optically coupled two internal states are dynamically entangled to form the Kondo-singlet state, overcoming the heating effect due to the irradiation. Furthermore, a lack of SU($N$) symmetry in the optical coupling provides a peculiar feature in the Kondo effect, which results in spin-selective renormalization of effective masses. We also discuss the effects of interorbital exchange interactions, and reveal that they induce novel crossover or reentrant behavior of the Kondo effect owing to control of the coupling anisotropy. The laser-induced Kondo effect is highly controllable by tuning the laser strength and the frequency, and thus offers a versatile platform to study the Kondo physics using ultracold atoms.
\end{abstract}

\pacs{67.85.-d, 37.10.Jk, 75.30.Mb}

\maketitle


Even after a long history, the Kondo effect is still one of the central problems in condensed matter physics since it captures many essential aspects of strongly correlated systems. This effect arises from the screening of a localized spin by a surrounding fermion cloud due to antiferromagnetic coupling, forming an emergent entangled state called the Kondo singlet \cite{Hewson}. 
In some rare-earth compounds where localized spins are periodically aligned, the Kondo effect takes place at every lattice site, and the quasiparticle band arising from the Kondo singlets forms a strongly renormalized heavy-fermion liquid which gives rise to diverse phenomena such as quantum criticality \cite{Colemanreview}. On the other hand, recent development of manipulation techniques using intense laser fields offers versatile methods to engineer intriguing states of quantum matter \cite{Jaksch, Dalibard}. Laser-dressed states are a useful resource both in electronic and cold-atomic systems \cite{Lindner, Eckstein, Takayoshi, Cooper, Jiang, Nakagawa}, but controlling the nature of these states in strongly correlated systems is challenging since we must deal with both strong-correlation effects and the influence of driving nonperturbatively \cite{Aoki}. The Kondo effect under irradiation provides a prototypical example of this problem. For example, in systems such as quantum dots, its nonequilibrium properties \cite{Heyl, Werner, Hettler, Nordlander} and optical phenomena in the Kondo effects \cite{Shahbazyan, Latta, Tureci, Sbierski} were investigated.
 
In this Letter, we propose a novel method to engineer the Kondo effect using intense laser fields and ultracold fermions in optical lattices. Our main idea is to realize the (effective) antiferromagnetic coupling between the localized spins and the fermion cloud using optical transitions induced by the laser. We show that the optical coupling dynamically entangles the spins with the cloud due to the Kondo effect, and thereby realizes a heavy-fermion liquid. A possible drawback in this scheme is the heating effect caused by the application of the laser. Here we find that the dynamically induced heavy-fermion liquid indeed persists under the irradiation, even if there is a substantial heating effect. Since the Kondo coupling can be tuned by the laser strength and frequency, the laser-induced Kondo effect offers a highly controllable scheme to generate the Kondo states in ultracold atoms. Moreover, it gives a natural platform to study real-time dynamics of the Kondo effect in various nonequilibrium situations.

As a suitable system to realize the Kondo effect using light-matter interactions, we focus on ultracold alkaline-earth-like atoms (AEAs) \cite{Fukuhara, Taie1, Taie2, Pagano, Cappellini, Scazza, DeSalvo, Martin, Zhang} in optical lattices. AEAs have an electronic ground state ($^1S_0$) and a long-lived metastable excited state ($^3P_0$). The stable and metastable states compose a two-orbital system, and the coherent optical process due to the optical transition with an ultranarrow linewidth plays an essential role for our purpose. Furthermore, another striking property of AEAs is that the spin degrees of freedom which come purely from the nuclear spin $I$ offer SU($N=2I+1$) spins with $N\leq 10$ \cite{Gorshkov}. It is shown that the interplay between the optical coupling and the multicomponent spin structure of AEAs leads to various unique features in the laser-induced Kondo effect, in contrast to previous proposals of the Kondo effect using ultracold atoms \cite{Gorshkov, Paredes, FossFeig1, FossFeig2, Bauer, Nishida}.

\begin{figure}
\includegraphics[width=8.5cm]{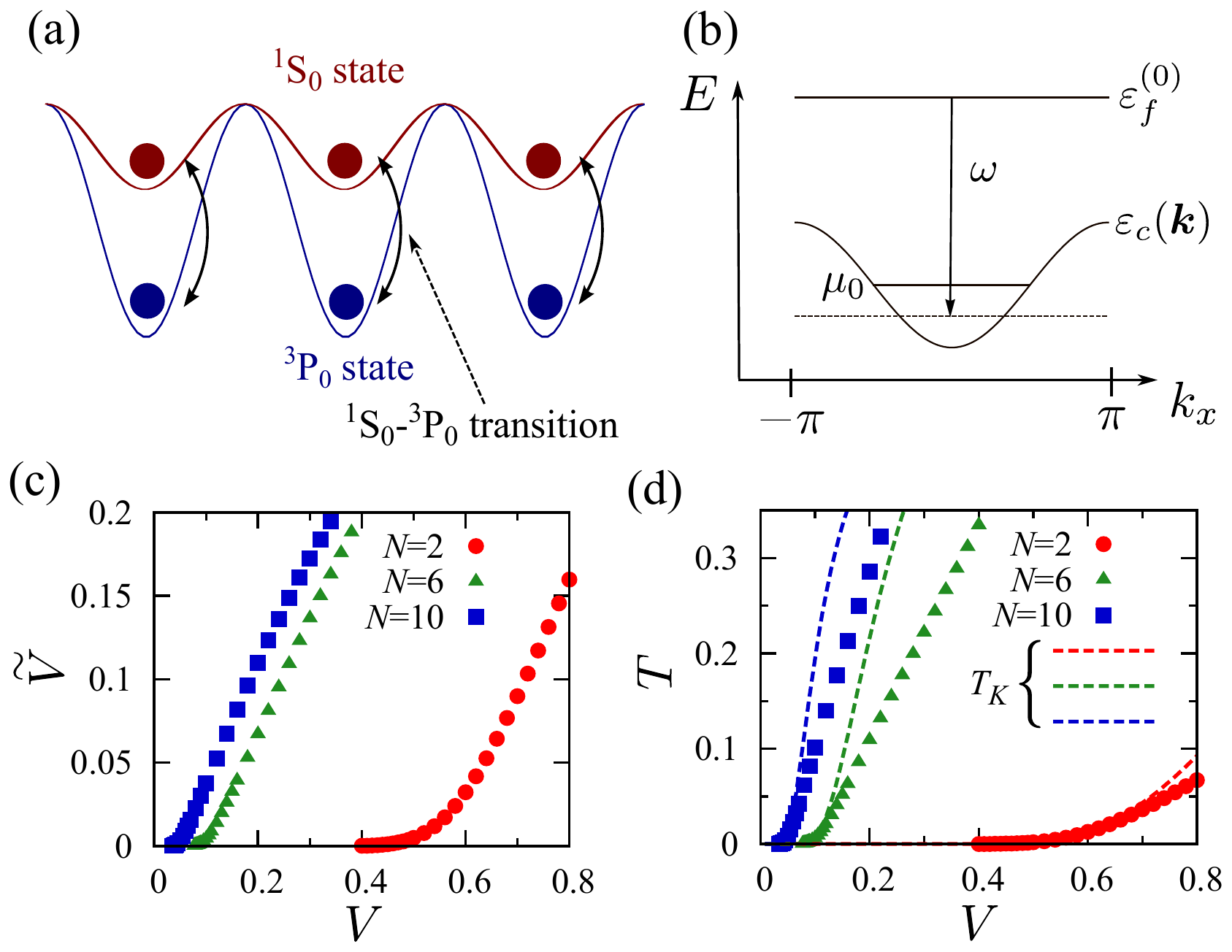}
\caption{(color online) (a) Schematic picture of the setup. Fermionic atoms in the $^1S_0$ state can move between sites of the optical lattice, while those in the $^3P_0$ state cannot. (b) The energy level shift after the gauge transformation. (c) Renormalized hybridization calculated with parameters $D=1, \varepsilon_f=-0.5, T_0=0, \mu_0=-0.1$, and $N=2, 6, 10$. (d) Effective temperatures calculated with the same parameters. The broken lines show the Kondo temperature $T_K$.}
\label{zerotemp}
\end{figure}

\textit{Setup}.--- We consider ultracold fermionic AEAs in a three-dimensional optical lattice which consist of the $^1S_0$ state and the $^3P_0$ state. The fermionic annihilation operators of the $^1S_0$ and $^3P_0$ states at site $i$ are denoted as $c_{i\sigma}$ and $f_{i\sigma}$ respectively, and also as $c_{\bm{k}\sigma}$ and $f_{\bm{k}\sigma}$ for their Fourier transform. Here $\sigma=-I, \cdots, I$ labels the $z$ component of the nuclear spin. The system is initially prepared with the Hamiltonian
\begin{align}
\mathcal{H}_{0}=\sum_{\bm{k},\sigma}(\varepsilon_c(\bm{k})c_{\bm{k}\sigma}^\dag c_{\bm{k}\sigma}
+\varepsilon_f^{(0)}f_{\bm{k}\sigma}^\dag  f_{\bm{k}\sigma})+U\sum_{i,\sigma<\sigma'}n_{fi\sigma}n_{fi\sigma'}
\end{align}
with $n_{fi\sigma}=f_{i\sigma}^\dag f_{i\sigma}$, where we denote the band dispersion of each orbital as $\varepsilon_c(\bm{k})$ and $\varepsilon_f^{(0)}$. The setup is illustrated in Fig. \ref{zerotemp} (a). Using the difference of polarizability of the two electronic states, an orbital-dependent optical lattice can be implemented in a manner such that the $^3P_0$ state feels a deep lattice potential while the $^1S_0$ state is confined weakly \cite{Gorshkov}. We assume that the $^3P_0$ state is completely localized due to the strong confinement and thus is dispersionless, experiencing strong on-site repulsion, $U>0$. For the moment, we neglect other interatomic interactions. In the initial setup, we assume that the system is in thermal equilibrium with temperature $T_0$. The chemical potential for the $^1S_0$ orbital is $\mu_0$, and the immobile $^3P_0$ state is assumed to be singly occupied at each site.

At time $t=0$, we apply a monochromatic laser which induces the $^1S_0$-$^3P_0$ transition allowed by the electric dipole coupling
\begin{gather}
\mathcal{H}_\mathrm{mix}=\theta(t)\sum_{i,\sigma,\sigma'}(\bm{V}\cdot\bm{\sigma}_{\sigma\sigma'}e^{i\bm{K}\cdot\bm{R}_i-i\omega t}f_{i\sigma}^\dag c_{i\sigma'}+\mathrm{H.c.})\label{Lmix}.
\end{gather}
Here we consider the simplest setup for the application of the laser, and $\theta(t)$ is the Heaviside step function. After $t=0$, the system evolves in time under the Hamiltonian $\mathcal{H}_0+\mathcal{H}_\mathrm{mix}$. The coupling coefficient in $\mathcal{H}_{\mathrm{mix}}$ follows from matrix elements of $-\bm{d}\cdot\bm{E}$, where $\bm{d}$ is the electric dipole moment and $\bm{E}$ is the electric field of light. Using the Wigner-Eckart theorem \cite{Landau}, the matrix elements of $\bm{d}$ are proportional to those of the nuclear spin operator $\bm{\sigma}_{\sigma\sigma'}$, since there is no electronic angular momentum.

Our central idea is that the Hamiltonian $\mathcal{H}=\mathcal{H}_0+\mathcal{H}_\mathrm{mix}$ mimics the Anderson lattice model for heavy-fermion systems, and the hybridization term $\mathcal{H}_{\mathrm{mix}}$ induces an effective antiferromagnetic interaction between the orbitals \cite{Hewson, Colemanreview}. After a gauge transformation $f_{i\sigma}(t)=e^{i\bm{K}\cdot\bm{R}_i-i\omega t}f_{i\sigma}'(t)$, the optical coupling takes a stepwise form in time. Hence, in this setup the hybridization term is suddenly turned on, and this corresponds to a "hybridization quench" problem of the Anderson lattice model. The transformation involves the energy level shift in the $^3P_0$ orbital, $\varepsilon_f\equiv\varepsilon_f^{(0)}-\omega$, as depicted in Fig. \ref{zerotemp} (b), which comes from the time-derivative term in the Lagrangian.

\textit{Emergence of the Kondo effect under the laser field}.--- In this Letter, we focus on properties of the steady state under the irradiation. To analyze the Kondo effect arising from the Hamiltonian $\mathcal{H}$, we employ slave-boson mean-field theory \cite{ReadNewns,Coleman,MillisLee}, which becomes exact in the $N\to\infty$ limit ($N=2I+1$ is the number of spin components). As the first approximation, we take the strong-correlation limit $U\to\infty$ and neglect the double occupancy of the $^3P_0$ state. Next, we split the $f$ operators into $f_{i\sigma}'=b_i^\dag \tilde{f}_{i\sigma}$ by introducing a slave-boson operator $b_i$, with a constraint $\sum_{\sigma}\tilde{f}_{i\sigma}^\dag \tilde{f}_{i\sigma}+b_i^\dag b_i=1$ at each site. To impose the constraint, the Lagrange multiplier field $\lambda_i$ is added to the Hamiltonian. After integrating out the fermions and taking a variation of the action with respect to $b_i$ and $\lambda_i$, we can obtain the saddle-point conditions in the (real-time) path-integral formalism (the derivation is given in \cite{supple}). For simplicity, we set a homogeneous ansatz for the mean-field solution, neglecting the effect of the trap potential to atoms. We obtain the saddle-point conditions for the steady state as
\begin{gather}
\tilde{V}(\tilde{\varepsilon}_f-\varepsilon_f)+\frac{V^2}{2N_s}\sum_{\bm{k},\sigma,\sigma'}(\hat{\bm{V}}\cdot\bm{\sigma}_{\sigma\sigma'})^*iG^K_{fc,\bm{k}\sigma\sigma'}(t,t)=0,\label{gapeq}\\
\frac{1}{N_s}\sum_{\bm{k},\sigma}iG^<_{ff,\bm{k}\sigma\sigma}(t,t)+\Bigr(\frac{\tilde{V}}{V}\Bigl)^2-1=0,\label{constraint}
\end{gather}
where $G^<_{ff,\bm{k}\sigma\sigma}(t,t')=i\langle\tilde{f}^\dag_{\bm{k}\sigma}(t)\tilde{f}_{\bm{k}\sigma}(t')\rangle$ and $G^K_{fc,\bm{k}\sigma\sigma'}(t,t')=-i\langle[\tilde{f}_{\bm{k}\sigma}(t),c^\dag_{\bm{k}\sigma'}(t')]\rangle$ are the lesser and Keldysh components of the Green's function, respectively \cite{Kamenev}. $N_s$ denotes the number of sites. The mean fields $b$ and $\lambda$ (here we have dropped the site indices) are incorporated in the renormalized hybridization $\tilde{\bm{V}}\equiv\bm{V}b/\sqrt{N_s}$ and the renormalized energy level, $\tilde{\varepsilon}_f\equiv\varepsilon_f+\lambda/\sqrt{N_s}$. These renormalization effects are the manifestation of the strong correlation arising from the Kondo effect. We set $V=|\bm{V}|, \tilde{V}=|\tilde{\bm{V}}|$, and $\hat{\bm{V}}=\bm{V}/|\bm{V}|$.

After the gauge transformation, the Hamiltonian at $t>0$ is time independent. In this Letter, instead of solving the full time-dependent mean-field theory, we simply assume that the steady state after a long time is a thermal equilibrium state described by an effective temperature $T$ and a chemical potential $\mu$. Since the external driving leads to heating of the system, the temperature $T$ and the chemical potential $\mu$ after the application of the laser are different from $T_0$ and $\mu_0$. The Kondo effect occurs below the characteristic temperature, the so-called Kondo temperature $T_K$. If the effective temperature $T$ exceeds the Kondo temperature, the Kondo effect is washed out by the heating. To address the effect of heating by the application of the laser, we evaluate $T$ and $\mu$ using the energy conservation law and the particle number conservation law (the explicit forms are found in \cite{supple}), since the atoms are trapped in vacuum and well isolated from the environment. The parameters to be determined are hence $(\tilde{V}, \tilde{\varepsilon}_f, T, \mu)$, and they are obtained self-consistently from the four equations, namely, the two saddle-point conditions, Eqs. (\ref{gapeq}) and (\ref{constraint}), and the two conservation laws.

\begin{figure}
\includegraphics[width=8.5cm]{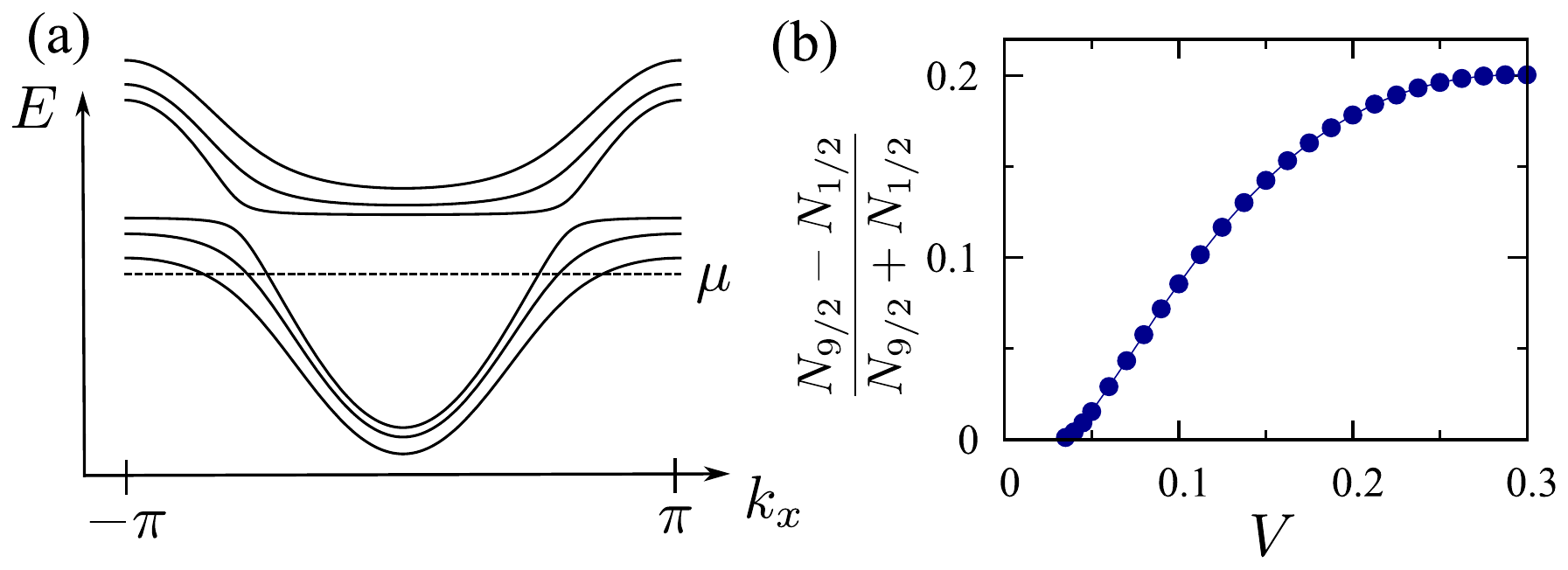}
\caption{(color online) (a) The renormalized quasiparticle band induced by the laser-induced Kondo effect for $N=6$. (b) Population imbalance induced by the Kondo effect. The $N=10$ case is shown, with parameters $D=1,\varepsilon_f=-0.8,T_0=0,\mu_0=-0.5$. $N_\sigma=n_\sigma+n_{-\sigma}$ is the particle number of each spin component.}
\label{popul_imbl}
\end{figure}

The self-consistent equations are solved numerically. In the numerical calculation, we assume that the laser has $\pi$ polarization $\bm{V}=(0,0,V)$, and that the density of states of the $^1S_0$ orbital is constant with finite bandwidth, $\rho(\varepsilon)=\rho_0=1/(2D), -D\leq\varepsilon\leq D$. The Kondo temperature $T_K$ is defined by the temperature where the slave boson starts to have a nonzero expectation value in thermal equilibrium \cite{Burdin}, as $\displaystyle T_K=c_N(D-\mu)^{1/N}(D+\mu)^{1-1/N}\exp \bigl[\frac{\varepsilon_f-\mu}{A_N\rho_0 V^2}\bigr]$ where $A_N=N(N^2-1)/12$ and $c_N$ is a weakly $N$-dependent constant of order one \cite{supple}. Figures \ref{zerotemp} (c) and \ref{zerotemp} (d) show the numerical solutions of the self-consistent equations starting from the zero-temperature initial state. The nonzero value of the renormalized hybridization, which entangles the localized $^3P_0$ atoms with the $^1S_0$ cloud, signals the emergence of the Kondo effect under the photoirradiation. The strongly renormalized value of $\tilde{V}$ leads to a hybridized band with a narrow Kondo gap (see Fig. \ref{popul_imbl} (a)), and thus realizes a heavy-fermion liquid. The effective temperature (Fig. \ref{zerotemp} (d)) is increased due to the heating effect caused by the irradiation. An important point is that \textit{the effective temperature is always lower than the Kondo temperature} $T_K$, and thus the Kondo effect can be realized. 
If the initial states are at finite temperatures, we can obtain similar results, although in this case the Kondo effect appears above some finite laser strength \cite{supple}. If the Fermi temperature is of the order of 100 nK, the necessary laser strength (or Rabi frequency) is estimated to be a few kHz, which is an achievable value in experiments.

Note that AEAs can offer a system with large-$N$ spin components, where $N=6$ for $^{173}$Yb \cite{Fukuhara, Taie1, Taie2, Pagano, Cappellini, Scazza} and $N=10$ for $^{87}$Sr \cite{DeSalvo, Martin, Zhang}. The Kondo temperature, which is the underlying energy scale of the Kondo physics, rapidly increases with the number of components $N$. This means that the Kondo state is more stable in large-$N$ systems, and makes the observation of the Kondo effect more feasible in the large spin state with AEAs which is experimentally realized \cite{Fukuhara, Taie1, Taie2, Pagano, Cappellini, Scazza, DeSalvo, Martin, Zhang, noteTK}.

We also comment on a role of the trap potential. Since the trap potential reduces $\varepsilon_f-\mu$, $T_K$ is lowest at the trap center, and there should appear a "mixed valence" region where $\varepsilon_f\lesssim\mu$ and the Kondo temperature is quite high. The existence of that region supports the feasibility to realize the Kondo liquid. However, we note that since the renormalization effect is weak in the mixed valence regime, a sharp Kondo effect may appear near the trap center.

\textit{Spin-selective renormalization}.--- The laser-induced Kondo state shows peculiar features owing to interplay between the optical coupling and the multicomponent spin structure of AEAs. In particular, the spin-dependent optical coupling leads to intriguing consequences with the Kondo state. The optical coupling for the $\pi$-polarized light becomes $\bm{V}\cdot\bm{\sigma}_{\sigma\sigma'}=\sigma V\delta_{\sigma\sigma'}$ and explicitly breaks the spin SU($N$) invariance. Therefore, the laser-driven Kondo gaps are spin dependent, as seen in the quasiparticle band structure $\varepsilon_{\sigma\pm}(\bm{k})=\frac{1}{2}(\varepsilon_c(\bm{k})+\tilde{\varepsilon}_f\pm\sqrt{(\varepsilon_c(\bm{k})-\tilde{\varepsilon}_f)^2+4|\sigma \tilde{V}|^2})$ (illustrated in Fig. \ref{popul_imbl} (a)) which is obtained by diagonalizing the mean-field Hamiltonian. This gives rise to spin-dependent effective masses of quasiparticles, since the quasiparticle band is more flattened in the higher spin components at the Fermi energy. As a result, the laser-induced Kondo effect selectively renormalizes the effective masses of higher spin components, beside light quasiparticles of lower spin components. These consequences are in marked contrast to the SU($N$)-symmetric Kondo effect proposed for ultracold atoms previously \cite{Gorshkov, Paredes, FossFeig1, FossFeig2, Bauer, Nishida}, where the all spin components are completely degenerate.

These characteristic spin-dependent structures are reflected in physical observables. The spin-dependent quasiparticle band yields population imbalance between the spin components. In Fig. \ref{popul_imbl} (b), we have plotted the population imbalance. The population imbalance, which can be directly measured by cold-atom experiments, inherits both the spin-dependent nature and nonperturbative $V$ dependence in the laser-induced Kondo effect. The Kondo states can also be measured by density profiles of the atomic cloud. For $N=2$, and when the $^1S_0$ orbital is half-filled in the lattice, the lower hybridized band formed by the Kondo effect is completely filled: this phase is called the Kondo insulator. Reflecting the formation of the Kondo insulator, a density plateau of the atomic cloud appears in some region of the atomic cloud \cite{FossFeig1, FossFeig2}. Similarly, in the case of $N>2$, reflecting the spin-dependent Kondo gaps, a half-metallic phase composed of a completely filled band (the Kondo insulator) and partially filled Kondo metals can be formed. Correspondingly, the density plateau may also be spin dependent for $N>2$.

\begin{figure}
\includegraphics[width=6cm]{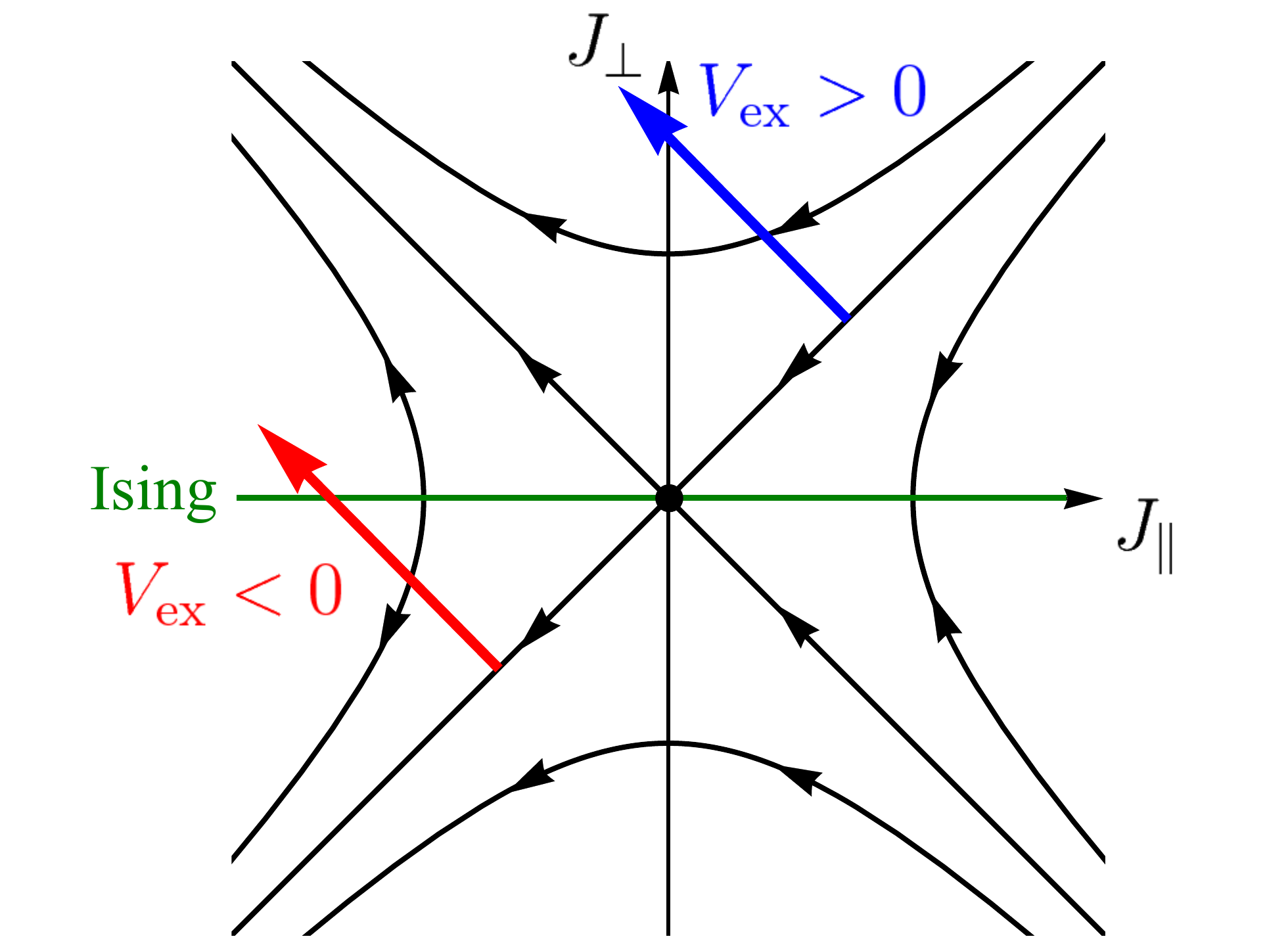}
\caption{(color online) The renormalization group flow for the case of the Kondo impurity for $N=2$. The blue (red) line shows a trajectory of the coupling constants when we apply the external laser in the case of $V_{\mathrm{ex}}>0$ ($V_{\mathrm{ex}}<0$).}
\label{RGflow}
\end{figure}

\textit{Effect of exchange interactions: Tuning anisotropy}.--- So far we have described the laser-induced Kondo effect using the minimal Hamiltonian $\mathcal{H}_0+\mathcal{H}_{\mathrm{mix}}$, which includes only interactions between the $^3P_0$ states. In reality, other interactions also exist \cite{Gorshkov}. Among them, strong interorbital SU($N$) exchange interactions observed in recent experiments \cite{Cappellini, Scazza, Zhang} play the most crucial role for the Kondo physics. Here we consider how these interactions affect the laser-induced Kondo effect. The SU($N$) exchange interaction is described by
\begin{equation}
\mathcal{H}_{\mathrm{ex}}=V_{\mathrm{ex}}\sum_{i,\sigma,\sigma'}c_{i\sigma}^{\dag}f_{i\sigma'}^{\dag}c_{i\sigma'}f_{i\sigma},
\end{equation}
where $V_{\mathrm{ex}}>0$ ($V_{\mathrm{ex}}<0$) is the ferromagnetic (antiferromagnetic) case. To gain insight into the Kondo effect in the case where both the exchange interaction and photo-induced hybridization exist, it is convenient to consider situations in which the laser frequency is highly off  and optical transitions are well suppressed. In this case, we can derive an effective Hamiltonian in the Hilbert subspace restricted to $\sum_\sigma n_{fi\sigma}=1$, using a standard treatment by the Schrieffer-Wolff transformation \cite{SW}. At the second order of $V$, we obtain an effective Kondo exchange interaction by the $\pi$-polarized laser as
\begin{equation}
\mathcal{H}_{\mathrm{eff}}=\sum_{i,\sigma,\sigma'}(V_{\mathrm{ex}}-J\sigma\sigma')c_{i\sigma}^{\dag}f_{i\sigma'}^{\dag}c_{i\sigma'}f_{i\sigma},\label{Heff}
\end{equation}
where $J\simeq 2V^2/|\varepsilon_f|>0$ \cite{potscatt}. The resulting exchange interaction, Eq. (\ref{Heff}), is anisotropic, 
and we find that the "tuning" of the anisotropy by the laser leads to an intriguing consequence. The effect of the anisotropic exchange coupling can be deduced from the renormalization group (RG) flow using Anderson's poor-man's scaling for the case of Kondo impurity, particularly in the case of $N=2$ \cite{Anderson}. 
The coupling constants 
$J_\perp=V_{\mathrm{ex}}+J/4$ and $J_\parallel=V_{\mathrm{ex}}-J/4$ 
change along a trajectory on the RG flow diagram as depicted in Fig. \ref{RGflow} when we increase $J$. 
When we turn on the laser in the case of $V_{\mathrm{ex}}>0$, the coupling constants immediately change from an irrelevant coupling into a relevant one; namely, the laser-induced Kondo effect occurs, despite the bare \textit{ferromagnetic} exchange coupling. This fact is in sharp contrast to the proposals in Refs. \cite{Gorshkov, FossFeig1, FossFeig2}. The behavior of the RG flow from the weak-coupling fixed point to the strong-coupling one may be observed as a crossover in the temperature dependence of the laser-induced Kondo effect.

In the case of $V_{\mathrm{ex}}<0$, where the bare exchange coupling is relevant, the Kondo effect owing to $V_{\mathrm{ex}}$ (as described in Refs. \cite{FossFeig1, FossFeig2}) takes place. This Kondo effect is suppressed by the laser application. At $J=4|V_{\mathrm{ex}}|$, the Kondo effect vanishes since the interaction is purely Ising-type, $J_\perp=0$. After that, as seen in the flow diagram, the Kondo effect revives and the system is governed by another strong-coupling fixed point which is described in this Letter. In other words, the laser gives a novel method to induce a reentrant Kondo effect across the Ising point.

For $N>2$, the situation is more complicated since more coupling constants appear in the Hamiltonian. We note that there is no "Ising" point where all spin-flip terms simultaneously vanish in this case, and thus the reentrant behavior of the Kondo effect may be most prominent in the case of $N=2$. A full analysis needs further investigations and is left for future work. 

\textit{Summary}.--- We have proposed a possible realization of the laser-induced Kondo state, using ultracold alkaline-earth atoms in an optical lattice. It has been shown that the emergent Kondo effect under the application of the laser field overwhelms the heating effect, thereby realizing the orbital-spin entangled singlets and the heavy-fermion liquid. Furthermore, we have elucidated peculiar Kondo physics arising from the specific form of optical coupling, such as the spin-selective renormalization of effective masses and the nontrivial competition between the bare exchange coupling and the laser-induced hybridization which leads to a novel crossover or reentrant behavior of the Kondo effect. 

The laser-induced Kondo effect proposed here provides a new avenue to study real-time dynamics of strongly correlated systems in nonequilibrium situations. One of interesting future perspectives is to address the transient process, where the Kondo singlet develops towards the steady state described in this Letter. Another interesting issue may be related to magnetic properties. Here we have only considered paramagnetic states, but Ruderman-Kittel-Kasuya-Yosida interactions mediated by the optical transitions are expected to cause various magnetic orders competing with the Kondo effect. This may lead to novel quantum criticality driven by external laser fields. These issues are expected to be clarified in future studies.

\begin{acknowledgments}
We are thankful to Yoshiro Takahashi for valuable discussions. This work was supported by KAKENHI (Grants No. 25400366, No. 14J01328, and No. 15H05855). M. N. thanks JSPS for the support from a Research Fellowship for Young Scientists.
\end{acknowledgments}

\bibliography{acKondo_ref.bib,acKondo_supple_ref.bib}

\begin{thebibliography}{46}
\bibitem [{not()}]{noteTK}%
  \BibitemOpen
  \href@noop {} {}\bibinfo {note} {For $N=2$, fluctuation effects beyond the
  mean-field theory are significant. It is known that the fluctuation effects
  reduce $T_K$ from the mean-field result. However, since the fluctuation also
  affects the energy spectrum, the effective temperature beyond the mean-field
  result is an interesting issue for a future study.}\BibitemShut {Stop}%
\bibitem [{pot()}]{potscatt}%
  \BibitemOpen
  \href@noop {} {}\bibinfo {note} {A potential scattering term from the
  Schrieffer-Wolff transformation can be absorbed into the chemical potential.
  We also remark that this transformation is not valid near the edge of the
  atomic cloud.}\BibitemShut {Stop}%
\end{thebibliography}

\pagebreak
\begin{widetext}
\section{Supplemental material}

\subsection{Derivation of the self-consistent equations}
We derive the saddle-point conditions using the real-time field theory, referred to as Keldysh formalism \cite{Kamenev}. Our action in the slave-boson representation reads
\begin{gather}
S=\int_C \mathrm{d}t (\mathcal{L}_{0}+\mathcal{L}_{\mathrm{mix}}),\notag\\
\mathcal{L}_{0}=\sum_{\bm{k},\sigma}c_{\bm{k}\sigma}^\dag(i\partial_t-\varepsilon_c(\bm{k}))c_{\bm{k}\sigma}
+\sum_{i,\sigma}\tilde{f}_{i\sigma}^\dag(i\partial_t-\varepsilon_f)\tilde{f}_{i\sigma}
+\sum_i b_i^\dag i\partial_t b_i-\sum_i\lambda_i\Bigl(\sum_{\sigma}\tilde{f}_{i\sigma}^\dag \tilde{f}_{i\sigma}+b_i^\dag b_i-1\Bigr),\notag\\
\mathcal{L}_{\mathrm{mix}}=-\sum_{i,\sigma,\sigma'}(\bm{V}\cdot\bm{\sigma}_{\sigma\sigma'}\tilde{f}_{i\sigma}^\dag b_i c_{i\sigma'}+\mathrm{h.c.}).\notag
\end{gather}
$C=C_+\cup C_-$ denotes the Keldysh contour \cite{Kamenev}, which is the sum of two contours from $t=0$ to $t=\infty$ and from $t=\infty$ to $t=0$. We denote a field on the contour $C_+$ and $C_-$ as $\varphi_+(t)$ and $\varphi_-(t)$, respectively ($\varphi=c,\tilde{f},b,\lambda$). Thanks to the doubling of the fields, this field theory contains full information of real-time correlators of the system, especially the Green's functions. For convenience, we introduce the Keldysh rotation of each field: for bosons,
\begin{gather}
\begin{pmatrix}
\chi_1(t) \\
\chi_2(t)
\end{pmatrix}
\equiv\frac{1}{\sqrt{2}}
\begin{pmatrix}
1 & 1 \\
1 & -1
\end{pmatrix}
\begin{pmatrix}
\chi_+(t) \\
\chi_-(t)
\end{pmatrix},\notag
\begin{pmatrix}
\chi_1^\dag(t) \\
\chi_2^\dag(t)
\end{pmatrix}
\equiv\frac{1}{\sqrt{2}}
\begin{pmatrix}
1 & 1 \\
1 & -1
\end{pmatrix}
\begin{pmatrix}
\chi_+^\dag(t) \\
\chi_-^\dag(t)
\end{pmatrix}\notag
\end{gather}
and for fermions,
\begin{gather}
\begin{pmatrix}
\psi_1(t) \\
\psi_2(t)
\end{pmatrix}
\equiv\frac{1}{\sqrt{2}}
\begin{pmatrix}
1 & 1 \\
1 & -1
\end{pmatrix}
\begin{pmatrix}
\psi_+(t) \\
\psi_-(t)
\end{pmatrix},\notag
\begin{pmatrix}
\psi_1^\dag(t) \\
\psi_2^\dag(t)
\end{pmatrix}
\equiv\frac{1}{\sqrt{2}}
\begin{pmatrix}
1 & -1 \\
1 & 1
\end{pmatrix}
\begin{pmatrix}
\psi_+^\dag(t) \\
\psi_-^\dag(t)
\end{pmatrix},\notag
\end{gather}
where the convention of Ref. \cite{Kamenev} is used. After the Keldysh rotation, the action turns into
\begin{gather}
S=S_c+S_f+S_b+S_{\mathrm{mix}}+\sqrt{2}\sum_{i}\lambda_{2i},\notag\\
S_c=\int_0^\infty\mathrm{d}t\sum_{\bm{k},\sigma}(c_{1\bm{k}\sigma}^\dag, c_{2\bm{k}\sigma}^\dag)
\begin{pmatrix}
i\partial_t-\varepsilon_c(\bm{k}) & 0 \\
0 & i\partial_t-\varepsilon_c(\bm{k})
\end{pmatrix}
\begin{pmatrix}
c_{1\bm{k}\sigma} \\
c_{2\bm{k}\sigma}
\end{pmatrix},\notag\\
S_f=\int_0^\infty\mathrm{d}t\sum_{i,\sigma}(\tilde{f}_{1i\sigma}^\dag, \tilde{f}_{2i\sigma}^\dag)
\begin{pmatrix}
i\partial_t-\varepsilon_f-\lambda_{1i}/\sqrt{2} & -\lambda_{2i}/\sqrt{2} \\
-\lambda_{2i}/\sqrt{2} & i\partial_t-\varepsilon_f-\lambda_{1i}/\sqrt{2}
\end{pmatrix}
\begin{pmatrix}
\tilde{f}_{1i\sigma} \\
\tilde{f}_{2i\sigma}
\end{pmatrix},\notag\\
S_b=\int_0^\infty\mathrm{d}t\sum_{i}(b_{1i}^\dag, b_{2i}^\dag)
\begin{pmatrix}
-\lambda_{2i}/\sqrt{2} & i\partial_t-\lambda_{1i}/\sqrt{2} \\
i\partial_t-\lambda_{1i}/\sqrt{2} & -\lambda_{2i}/\sqrt{2}
\end{pmatrix}
\begin{pmatrix}
b_{1i} \\
b_{2i}
\end{pmatrix},\notag\\
S_{\mathrm{mix}}=\int_0^\infty\mathrm{d}t\frac{-1}{\sqrt{2}}\sum_{i,\sigma,\sigma'}
\Bigl\{ \bm{V}\cdot\bm{\sigma}_{\sigma\sigma'} \tilde{f}_{1i\sigma}^\dag (b_{1i}c_{1i\sigma'}+b_{2i}c_{2i\sigma'})
+\bm{V}\cdot\bm{\sigma}_{\sigma\sigma'} \tilde{f}_{2i\sigma}^\dag (b_{2i}c_{1i\sigma'}+b_{1i}c_{2i\sigma'})
+\mathrm{h.c.}\Bigr\}.\notag
\end{gather}
The partition function is
\begin{align}
Z&=\int\mathcal{D}[c,\tilde{f},b,\lambda]e^{iS}\notag\\
&\equiv\int\mathcal{D}[b,\lambda]e^{iS_{\mathrm{eff}}}\notag.
\end{align}
In the second line, we integrate out the fermions and define the effective bosonic action $S_{\mathrm{eff}}$. The saddle-point condition of the path integral is obtained by
\begin{align}
\frac{\delta S_{\mathrm{eff}}}{\delta b_{\alpha i}(t)}=\Bigl\langle\frac{\delta S}{\delta b_{\alpha i}(t)}\Bigr\rangle=0,\notag\\
\frac{\delta S_{\mathrm{eff}}}{\delta \lambda_{\alpha i}(t)}=\Bigl\langle\frac{\delta S}{\delta \lambda_{\alpha i}(t)}\Bigr\rangle=0\notag
\end{align}
with $\alpha=1,2$, where $\langle\cdots\rangle$ is the expectation value with fixed $b_{\alpha i}$ and $\lambda_{\alpha i}$. These lead to
\begin{gather}
(i\partial_t-\frac{\lambda_{1i}}{\sqrt{2}})b_{2i}(t)-\frac{\lambda_{2i}}{\sqrt{2}}b_{1i}(t)
-\frac{1}{\sqrt{2}}\sum_{i,\sigma,\sigma'}(\bm{V}\cdot\bm{\sigma}_{\sigma\sigma'})^*(\langle c_{1i\sigma'}^\dag(t)\tilde{f}_{1i\sigma}(t)\rangle+\langle c_{2i\sigma'}^\dag(t)\tilde{f}_{2i\sigma}(t)\rangle)=0,\notag\\
(i\partial_t-\frac{\lambda_{1i}}{\sqrt{2}})b_{1i}(t)-\frac{\lambda_{2i}}{\sqrt{2}}b_{2i}(t)
-\frac{1}{\sqrt{2}}\sum_{i,\sigma,\sigma'}(\bm{V}\cdot\bm{\sigma}_{\sigma\sigma'})^*(\langle c_{2i\sigma'}^\dag(t)\tilde{f}_{1i\sigma}(t)\rangle+\langle c_{1i\sigma'}^\dag(t)\tilde{f}_{2i\sigma}(t)\rangle)=0,\notag\\
\sum_{\sigma}(\langle \tilde{f}_{1i\sigma}^\dag(t)\tilde{f}_{1i\sigma}(t)\rangle+\langle \tilde{f}_{2i\sigma}^\dag(t)\tilde{f}_{2i\sigma}(t)\rangle)
+b_{1i}^\dag(t)b_{2i}(t)+b_{2i}^\dag(t)b_{1i}(t)=0,\notag\\
\sum_{\sigma}(\langle \tilde{f}_{1i\sigma}^\dag(t)\tilde{f}_{2i\sigma}(t)\rangle+\langle \tilde{f}_{2i\sigma}^\dag(t)\tilde{f}_{1i\sigma}(t)\rangle)
+b_{1i}^\dag(t)b_{1i}(t)+b_{2i}^\dag(t)b_{2i}(t)-2=0.\notag
\end{gather}
When we focus on the homogeneous steady state, the set of saddle-point conditions can be simplified as
\begin{gather}
\lambda_1 b_2+\lambda_2 b_1+\frac{1}{N_s}\sum_{\bm{k},\sigma,\sigma'}(\bm{V}\cdot\bm{\sigma}_{\sigma\sigma'})^*(\langle c_{1\bm{k}\sigma'}^\dag(t)\tilde{f}_{1\bm{k}\sigma}(t)\rangle+\langle c_{2\bm{k}\sigma'}^\dag(t)\tilde{f}_{2\bm{k}\sigma}(t)\rangle)=0,\label{NESS1}\\
\lambda_1 b_1+\lambda_2 b_2+\frac{1}{N_s}\sum_{\bm{k},\sigma,\sigma'}(\bm{V}\cdot\bm{\sigma}_{\sigma\sigma'})^*(\langle c_{2\bm{k}\sigma'}^\dag(t)\tilde{f}_{1\bm{k}\sigma}(t)\rangle+\langle c_{1\bm{k}\sigma'}^\dag(t)\tilde{f}_{2\bm{q}\sigma}(t)\rangle)=0,\\
\frac{1}{N_s}\sum_{\bm{k},\sigma}(\langle \tilde{f}_{1\bm{k}\sigma}^\dag(t)\tilde{f}_{1\bm{k}\sigma}(t)\rangle+\langle \tilde{f}_{2\bm{k}\sigma}^\dag(t)\tilde{f}_{2\bm{k}\sigma}(t)\rangle)+b_1^\dag b_2+b_2^\dag b_1=0,\label{NESS3}\\
\frac{1}{N_s}\sum_{\bm{k},\sigma}(\langle \tilde{f}_{1\bm{k}\sigma}^\dag(t)\tilde{f}_{2\bm{k}\sigma}(t)\rangle+\langle \tilde{f}_{2\bm{k}\sigma}^\dag(t)\tilde{f}_{1\bm{k}\sigma}(t)\rangle)+b_1^\dag b_1+b_2^\dag b_2-2=0.
\end{gather}
We note that Eqs. (\ref{NESS1}) and (\ref{NESS3}) are identically satisfied when we set $b_2=\lambda_2=0$ because of the relation between the retarded and advanced Green's function: $G^R(t,t)+G^A(t,t)=0$ (See also Ref. \cite{Ratiani}).
Hence, we arrive at the saddle-point conditions for the steady state:
\begin{gather}
-\lambda_1 b_1-\sum_{\bm{k},\sigma,\sigma'}(\bm{V}\cdot\bm{\sigma}_{\sigma\sigma'})^*\langle c_{2\bm{k}\sigma'}^\dag(t)\tilde{f}_{1\bm{k}\sigma}(t)\rangle=0,\label{SC1}\\
\sum_{\bm{k},\sigma}\langle \tilde{f}_{2\bm{k}\sigma}^\dag(t)\tilde{f}_{1\bm{k}\sigma}(t)\rangle
+b_1^\dag b_1-2N_s=0.
\label{SC2}
\end{gather}
Using the relation $\langle c_{2\bm{k}\sigma'}^\dag(t)\tilde{f}_{1\bm{k}\sigma}(t)\rangle=iG^K_{fc,\bm{k}\sigma\sigma'}(t,t)$ and $\langle \tilde{f}_{2\bm{k}\sigma}^\dag(t)\tilde{f}_{1\bm{k}\sigma}(t)\rangle=1-iG^K_{ff,\bm{k}\sigma\sigma}(t,t)=2iG^<_{ff,\bm{k}\sigma\sigma}(t,t)$, we obtain the saddle-point conditions in the main text.

The rest of self-consistent equations consists of the energy conservation and the particle number conservation. These read
\begin{align}
&\sum_{\bm{k},\sigma}(\varepsilon_c(\bm{k})\langle c_{\bm{k}\sigma}^\dag c_{\bm{k}\sigma}\rangle_{\mathrm{ini}}+\varepsilon_f\langle f_{\bm{k}\sigma}^\dag f_{\bm{k}\sigma}\rangle_{\mathrm{ini}})\notag\\
=&\sum_{\bm{k},\sigma}(\varepsilon_c(\bm{k})\langle c_{\bm{k}\sigma}^\dag(t) c_{\bm{k}\sigma}(t)\rangle+\tilde{\varepsilon}_f\langle \tilde{f}_{\bm{k}\sigma}^\dag(t) \tilde{f}_{\bm{k}\sigma}(t)\rangle)
+\sum_{\bm{k},\sigma,\sigma}(\tilde{\bm{V}}\cdot\bm{\sigma}_{\sigma\sigma'}\langle \tilde{f}_{\bm{k}\sigma}^\dag(t) c_{\bm{k}\sigma'}(t)\rangle+\mathrm{h.c.})+\sum_i\lambda_i(|b_i|^2-1)\label{SC3},
\end{align}
\begin{align}
\sum_{\bm{k},\sigma}(\langle c_{\bm{k}\sigma}^\dag c_{\bm{k}\sigma}\rangle_{\mathrm{ini}}+\langle f_{\bm{k}\sigma}^\dag f_{\bm{k}\sigma}\rangle_{\mathrm{ini}})
=\sum_{\bm{k},\sigma}(\langle c_{\bm{k}\sigma}^\dag(t) c_{\bm{k}\sigma}(t)\rangle+\langle f_{\bm{k}\sigma}^\dag(t) f_{\bm{k}\sigma}(t)\rangle)\label{SC4},
\end{align}
where $\langle\cdots\rangle_{\mathrm{ini}}$ is the expectation value for the initial state, namely $\langle c_{\bm{k}\sigma}^\dag c_{\bm{k}\sigma}\rangle_{\mathrm{ini}}=f(\varepsilon_c(\bm{k}))$ and $\sum_{\bm{k},\sigma}\langle f_{\bm{k}\sigma}^\dag f_{\bm{k}\sigma}\rangle_{\mathrm{ini}}=1$ ($f(\varepsilon)$ is the Fermi distribution function).

To calculate the mean-field equations, we use the Green's functions of the mean-field Hamiltonian. For $\pi$-polarized laser, these reads 
\begin{align}
G^<_{cc,\bm{k}\sigma\sigma}(t,t')&=i\langle c_{\bm{k}\sigma}^\dag(t')c_{\bm{k}\sigma}(t)\rangle\notag\\
&=i|u_{\bm{k}\sigma}|^2f(\varepsilon_{\sigma+}(\bm{k}))e^{-i\varepsilon_{\sigma+}(\bm{k})(t-t')}+i|v_{\bm{k}\sigma}|^2f(\varepsilon_{\sigma-}(\bm{k}))e^{-i\varepsilon_{\sigma-}(\bm{k})(t-t')}\notag\\
G^<_{ff,\bm{k}\sigma\sigma}(t,t')&=i\langle \tilde{f}_{\bm{k}\sigma}^\dag(t')\tilde{f}_{\bm{k}\sigma}(t)\rangle\notag\\
&=i|v_{\bm{k}\sigma}|^2f(\varepsilon_{\sigma+}(\bm{k}))e^{-i\varepsilon_{\sigma+}(\bm{k})(t-t')}+i|u_{\bm{k}\sigma}|^2f(\varepsilon_{\sigma-}(\bm{k}))e^{-i\varepsilon_{\sigma-}(\bm{k})(t-t')},\notag\\
G^<_{fc,\bm{k}\sigma\sigma}(t,t')&=i\langle c_{\bm{k}\sigma}^\dag(t')\tilde{f}_{\bm{k}\sigma}(t)\rangle\notag\\
&=iu_{\bm{k}\sigma}v_{\bm{k}\sigma}^*f(\varepsilon_{\sigma+}(\bm{k}))e^{-i\varepsilon_{\sigma+}(\bm{k})(t-t')}-iu_{\bm{k}\sigma}v_{\bm{k}\sigma}^*f(\varepsilon_{\sigma-}(\bm{k}))e^{-i\varepsilon_{\sigma-}(\bm{k})(t-t')},\notag
\end{align}
and
\begin{align}
G^K_{fc,\bm{k}\sigma\sigma}(t,t')=-&iu_{\bm{k}\sigma}v_{\bm{k}\sigma}^*(1-2f(\varepsilon_{\sigma+}(\bm{k})))e^{-i\varepsilon_{\sigma+}(\bm{k})(t-t')}\notag\\
&+iu_{\bm{k}\sigma}v_{\bm{k}\sigma}^*(1-2f(\varepsilon_{\sigma-}(\bm{k})))e^{-i\varepsilon_{\sigma-}(\bm{k})(t-t')},\notag
\end{align}
where
\begin{align}
&u_{\bm{k}\sigma}=\sqrt{\frac{1}{2}\Bigl(1+\frac{\varepsilon_c(\bm{k})-\tilde{\varepsilon}_f}{E_\sigma(\bm{k})}\Bigr)},
v_{\bm{k}\sigma}=\sqrt{\frac{1}{2}\Bigl(1-\frac{\varepsilon_c(\bm{k})-\tilde{\varepsilon}_f}{E_\sigma(\bm{k})}\Bigr)},\notag\\
&\varepsilon_{\sigma\pm}(\bm{k})=\frac{\varepsilon_c(\bm{k})-\tilde{\varepsilon}_f\pm E_\sigma(\bm{k})}{2},\notag\\
&E_\sigma(\bm{k})=\sqrt{(\varepsilon_c(\bm{k})-\tilde{\varepsilon}_f)^2+4|\sigma\tilde{V}|^2}.\notag
\end{align}

In equilibrium states, the saddle point conditions (\ref{SC1}) and (\ref{SC2}) have the same form as that derived using standard Matsubara formalism \cite{Coleman, MillisLee}. Thus the Kondo temperature is obtained by setting $b_1=0$ and $T=T_K$ in Eqs. (\ref{SC1}) and (\ref{SC2}). First, using Eq. (\ref{SC2}), we obtain
\begin{equation*}
\tilde{\varepsilon}_f|_{T=T_K}=\mu+T_K\log (N-1)
\end{equation*}
because of the condition $Nf(\tilde{\varepsilon}_f)=1$. Substituting this into Eq. (\ref{SC1}), and using the constant density of states as mentioned in the main text, we arrive at
\begin{gather}
T_K=c_N(D-\mu)^{1/N}(D+\mu)^{1-1/N}\exp \bigl[\frac{\varepsilon_f-\mu}{A_N\rho_0 V^2}\bigr],\notag
\end{gather}
where
\begin{gather}
A_N\equiv \frac{N(N^2-1)}{12},\notag\\
\log c_N\equiv -(N-1)\int_0^\infty \mathrm{d}x\log x\cdot\Bigl\{\frac{e^x}{((N-1)e^x+1)^2}+\frac{e^{-x}}{((N-1)e^{-x}+1)^2}\Bigr\}.\notag
\end{gather}

\subsection{Numerical solutions in the case of finite initial temperatures}

In this section, we show numerical results for the case where the initial states are at finite temperatures. Fig. \ref{finitetemp} shows the numerical solutions of the mean-field equations at finite initial temperatures. In this case, the laser-induced Kondo effect emerges above some threshold value of laser strength, since the Kondo effect cannot appear in the region where $T_K<T_0$. When the optical coupling is weak, the photo-induced hybridization between the orbitals is washed out due to the thermal fluctuations. With sufficiently strong laser fields, the two orbitals start to entangle and show the Kondo effect. As we increase the laser strength, the thermal fluctuations become less important, and the solutions approach the case of zero initial temperatures described in the main text.

\begin{figure}[h]
\includegraphics[width=12cm]{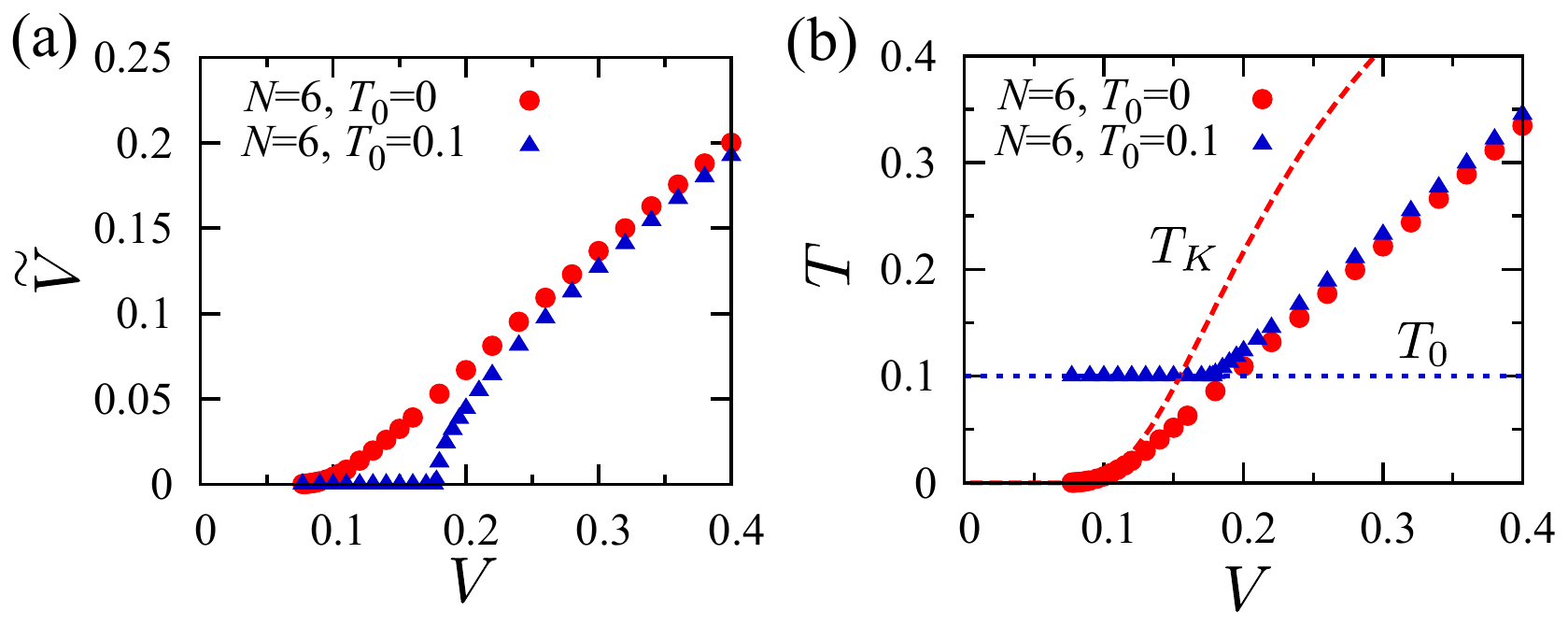}
\caption{(color online) Solutions of the mean-field equations starting from finite initial temperatures, with parameters $D=1, \varepsilon_f=-0.5, T_0=0.1, \mu_0=-0.1$, and $N=6$. The solutions at zero initial temperature is also shown for comparison. (a) Renormalized hybridization. (b) Effective temperature. The broken line shows the Kondo temperature, and the horizontal dotted line is the initial temperature.}
\label{finitetemp}
\end{figure}


\end{widetext}

\end{document}